\documentclass{article}

\usepackage{microtype}
\usepackage{graphicx}
\usepackage{subcaption}
\usepackage{booktabs} 

\usepackage{hyperref}

\usepackage[accepted]{icml2026}

\usepackage{amsmath}
\usepackage{amssymb}
\usepackage{mathtools}
\usepackage{amsthm}
\usepackage{amsfonts}

\usepackage[capitalize,noabbrev]{cleveref}

\theoremstyle{plain}

\theoremstyle{definition}

\theoremstyle{remark}

\usepackage[textsize=tiny]{todonotes}

\icmltitlerunning{HERDS: Strategic Defense Under Partial Observability with Semi-Bandit Feedback}

\begin{document}

\twocolumn[
  \icmltitle{Online Learning of Strategic Defense against Ecological Adversaries under Partial Observability with Semi-Bandit Feedback}


  \icmlsetsymbol{equal}{*}

  \begin{icmlauthorlist}
    \icmlauthor{Anjali Purathekandy}{a}
    \icmlauthor{Deepak N. Subramani}{a}
  \end{icmlauthorlist}

  \icmlaffiliation{a}{Department of Computational and Data Sciences, Indian Institute of Science, Bangalore, India}

  \icmlcorrespondingauthor{Deepak N. Subramani}{deepakns@iisc.ac.in}

  \icmlkeywords{Online Learning, Game Theory, Green Security Games, Agent-Based Modelling, Human-Elephant Conflicts}

  \vskip 0.3in
]



\printAffiliationsAndNotice{}  

\begin{abstract}
  We introduce an online learning algorithm for computing adaptive resource allocation policies against strategic ecological adversaries with unknown behavioral models and partial observability. Our setting addresses a fundamental limitation of security games: when adversary behavior cannot be modeled a priori, classical equilibrium-based approaches fail. We formulate the problem as regret minimization in a combinatorial action space with semi-bandit feedback, where payoffs are non-stationary and interdependent across targets. Our algorithm, named HERDS (Human-Elephant conflict mitigation through Resource Deployment for Strategic guarding), extends Follow-the-Perturbed-Leader (FPL) with three innovations: (1) simultaneous exploration-exploitation through dynamic budget partitioning driven by observed losses, (2) adaptive payoff estimation under confounded observations where attack entry points are unidentifiable, and (3) model-agnostic learning that provides regret guarantees without behavioral assumptions. We demonstrate our framework on Human-Elephant Conflict mitigation, a domain where intelligent ecological adversaries exhibit strategic behavior (optimal foraging, spatial memory, adaptive evasion) yet lack tractable behavioral models. Experiments using an Agent-Based Model calibrated with elephant movement data demonstrate 15--45\% regret reduction versus Follow-the-Perturbed-Leader with Uniform-Exploration (FPL-UE), 40--50\% crop damage reduction against adaptive adversaries, and convergence in 40--50 rounds versus 60--80 for baselines. 
\end{abstract}

\section{Introduction}

Human-Elephant Conflict (HEC) results in approximately 400 human deaths and 100 elephant deaths annually in India alone, directly impacting 500,000 families through crop raiding \cite{Fernando2008,Shaffer2019}. While communities employ various deterrents (noise, fire, barriers) and active guarding, elephants' high intelligence, spatial memory, and adaptability cause most mitigation strategies to lose effectiveness over time \cite{Fernando2008}. Implementing organized crop protection is costly, requiring trained personnel, equipment, and recurring budgets.

Decision analysis and game theory have proven to be valuable in biodiversity management and conservation \cite{Frank2010}. Field experience shows that elephants function as strategic adversaries with unknown behavioral models. They exhibit optimal foraging (targeting high-value crops \cite{Shaffer2019}), risk minimization (nocturnal raiding \cite{Fernando2008}), spatial memory, and adaptive learning to exploit security gaps. However, in contrast to standard security games where adversary behavior can be inferred from data, HEC presents a fundamental paradox: we observe clear evidence of strategic conduct in the field, yet we lack any behavioral models of elephants’ crop-raiding choices and do not have enough data on elephant–guard interactions to build such models. Substantial behavioral variation across locations and rapidly shifting conditions further undermine the feasibility of conventional methods.

This problem calls for an online learning approach. Strategic adversaries adjust their tactics, so fixed policies fail as elephants adapt to predictable patrol routes. Bayesian and model-based methods depend on prior distributions that we cannot credibly specify, given the lack of behavioral models and strong heterogeneity. In contrast, adversarial online learning offers regret guarantees against any adversary strategy without relying on accurate behavioral assumptions. It can also leverage historical data for initialization, while adaptively directing exploration efforts in response to observed crop damage. To develop and evaluate online learning strategies with security games, Agent-Based Models (ABMs) offer a safe virtual environment for testing policies when field experiments are risky \cite{Adam2011,Kirkland2020,Janssen2020}.

Field evidence indicates that guarding forest-agricultural boundaries is among the most effective mitigation approaches \cite{Fernando2008}. However, current guarding relies on uncoordinated individual efforts, resulting in inefficient resource allocation with redundant deployments and critical security gaps that intelligent elephants exploit. This creates a combinatorial optimization challenge: strategically allocate $K$ guards across $N$ boundary segments to minimize crop loss against adaptive adversaries who learn and respond to deployment patterns.

\begin{figure}[h]
  \centering
  \includegraphics[width=\linewidth]{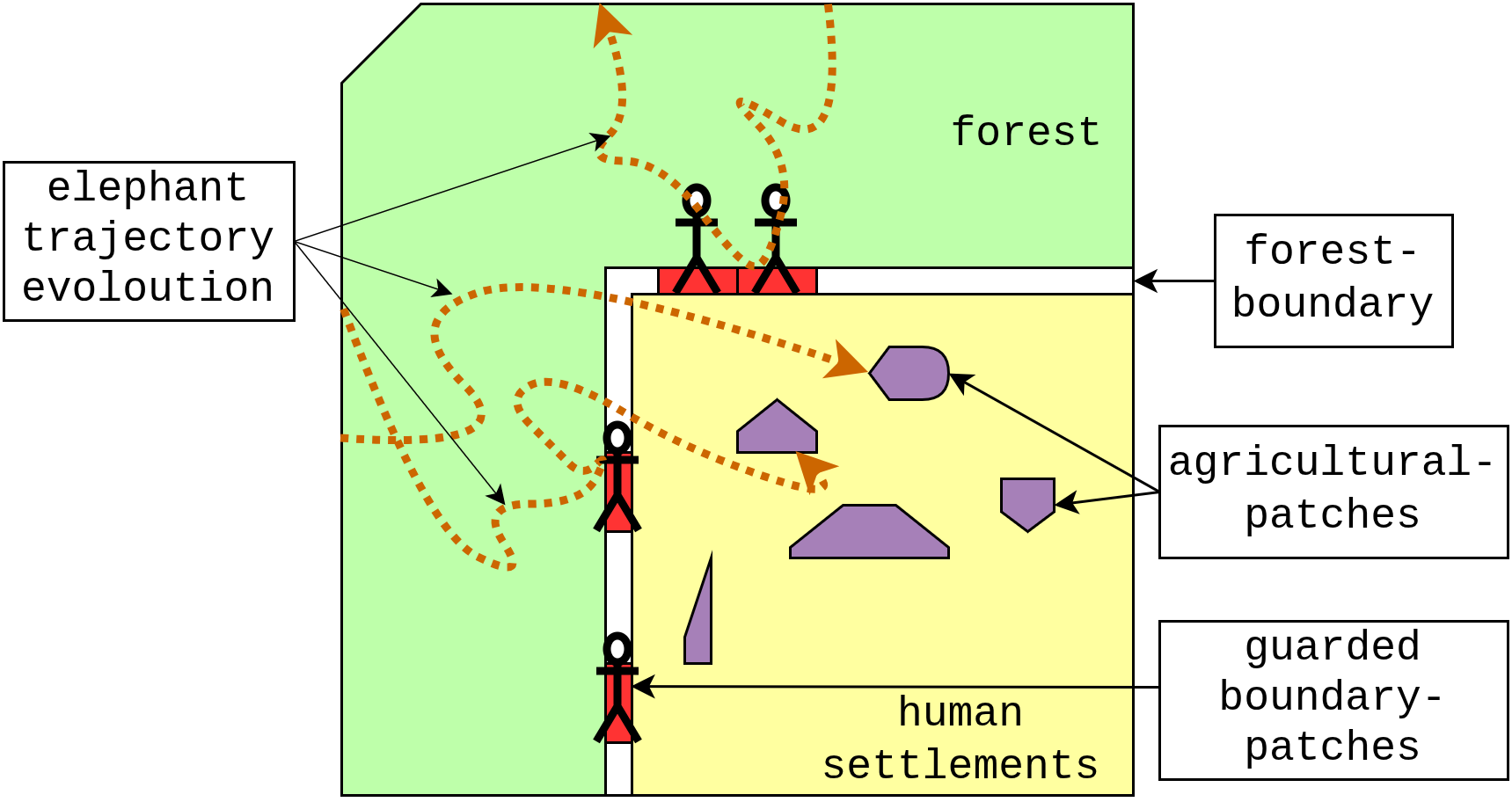}
  \caption{Schematic illustration of strategic guard placement along the forest-agricultural boundary to maximize interceptions. Three elephant trajectory scenarios are shown: (1) successful entry to the settlement via an unguarded boundary segment, resulting in crop raiding; (2) effective interception at a guarded segment, turning the elephant back into the forest; and (3) settlement entry despite ranger presence by using a different unguarded segment. Yellow areas depict human settlements and agricultural plots; white areas indicate forest boundaries; green areas indicate forest; red symbols mark guard positions; and brown lines represent potential elephant paths.}
  \label{fig:schematic-attacker-trajectory-evolution}
\end{figure}

We cast HEC mitigation as a Green Security Game, where elephants are modeled as strategic adversaries and defenders allocate limited patrol resources along forest–agriculture boundaries (Figure~\ref{fig:schematic-attacker-trajectory-evolution}). This constitutes the first game-theoretic treatment of HEC, broadening security games to encompass ecological adversaries that exhibit adaptive strategic behavior. Since no reliable behavioral models exist, an online learning approach is required. To address this, we introduce HERDS (Human-Elephant conflict mitigation through Resource Deployment for Strategic guarding), which builds on the FPL-UE framework \cite{Haifeng2016} with three innovations: (1) dynamically adjusting the exploration–exploitation budget in response to observed crop damage, (2) adaptively learning payoffs under partial observability while accounting for interdependent boundary segments, and (3) jointly deploying exploration and exploitation patrols. HERDS is model-agnostic, inferring effective policies without assuming a specific behavioral model, and offers performance guarantees through regret minimization.

Our contributions are: \textit{(1)} First game-theoretic framework for HEC, modeling elephants as strategic adversaries within a Green Security Game (GSG) structure and extending security game theory to ecological domains with unknown behavioral models. \textit{(2)} Dynamic payoff learning mechanism updating estimates in real-time despite partial observability constraints. \textit{(3)} HERDS algorithm adapting online learning methodology for strategic games with unknown adversaries. \textit{(4)} Validation using an ABM calibrated with elephant data from India \cite{Purathekandy2024}, demonstrating superior performance against multiple adversarial models. 

\section{Related Work}

HEC mitigation integrates security game theory (modeling strategic interactions) with online learning (handling unknown adversary models). Related work spans three areas:

\textbf{Online Learning Foundations.} Online learning addresses sequential decision-making without complete environment knowledge \cite{Kalai2005,CesaBianchi2006}. Foundational work establishes regret minimization as the central performance criterion for adversarial environments. Multi-armed bandit algorithms provide sublinear regret bounds for exploration-exploitation tradeoffs \cite{Auer2002}, extended to continuous action spaces \cite{Awerbuch2008}. Partial monitoring frameworks address incomplete feedback \cite{Audibert2010}, while online geometric optimization tackles adaptive adversaries \cite{McMahan2004}.

\textbf{Online Learning for Security Games.} When adversary behavioral models are unknown, online learning provides adaptive solutions without prior assumptions. The FPL algorithm with Geometric Resampling addresses online combinatorial optimization under semi-bandit feedback \cite{Neu2013}. FPL-UE extends this with uniform exploration, achieving provably sublinear regret without prior knowledge of payoffs or attacker behavior \cite{Haifeng2016}. Critically, regret bounds hold regardless of adversary strategy, making it suitable for HEC where elephant behavior cannot be reliably modeled a priori. Recent work introduced reallocation time criteria alongside regret \cite{Zhu2023} and methods for computing multi-round plans \cite{Marecki2012}.

\textbf{Green Security Games: Framework Foundation.} Our work extends GSGs for wildlife protection. GSGs model defenders allocating limited resources against adversaries targeting valuable assets \cite{Fang2015,Fang2016}. Early algorithms (ASAP \cite{Paruchuri2007}) and large-scale applications (GUARDS, PROTECT \cite{An2011,Tambe2011}) demonstrated practical value. Wildlife applications include PAWS with poacher behavioral models \cite{Yang2014} and GMOP using POMDP formulations \cite{Qian2014}. However, traditional GSGs assume known adversary models to compute equilibria. HEC presents a critical distinction: elephants are strategic adversaries (optimal foraging, spatial memory, adaptive evasion) yet we lack behavioral models. We extend the GSG framework to handle strategic ecological adversaries with unknown models via online learning rather than equilibrium computation.

\textbf{Adaptive Learning in Wildlife Security.} Prior work learns adversary behavioral models during deployment: online algorithms for patrol allocation \cite{Richard2014}, SHARP for adversary adaptation \cite{Kar2015}, MINION-sm building implicit models \cite{Shahrzad2019}, and LIZARD using multi-armed bandits \cite{xu2021}. HERDS differs by learning defender policies directly without constructing explicit adversary models, providing model-agnostic robustness.

\section{Problem Formulation}

\subsection{Problem Overview}

We formulate HEC mitigation as an online combinatorial optimization problem. The forest-agricultural boundary is partitioned into $N$ discrete segments. A defender allocates $K$ guards ($K \ll N$) across these segments over $T$ rounds to minimize cumulative crop damage caused by elephant incursions. The adversary (elephant) selects entry points based on an unknown behavioral model, and the defender receives only partial feedback: interceptions are observed at guarded segments, but the specific entry points used at unguarded segments remain unknown. The objective is to learn an adaptive guard deployment policy that minimizes regret relative to the best fixed hindsight strategy.

\subsection{Key Departures from Classical Security Games} \label{sec:challenge}

Our formulation differs fundamentally from classical GSGs in three aspects that necessitate a new algorithmic approach:

\textbf{Challenge 1: Unknown Adversary Behavior.} Classical GSGs assume adversary response models can be estimated from historical data or domain knowledge, enabling equilibrium computation \cite{Fang2016}. In HEC, elephant behavioral strategies remain unknown. We observe strategic behavior (optimal foraging, adaptive evasion) but lack models of how elephants respond to guard deployments. The defender must learn effective policies without any prior behavioral assumptions.

\textbf{Challenge 2: Partial Observability with Confounded Feedback.} Standard security games assume that the defender observes attack outcomes at all targets. In HEC, the defender only observes interceptions in the guarded segments. When elephants successfully raid crops, the total damage is observable, but the specific boundary segment used for entry cannot be determined due to limited surveillance. This creates \textit{confounded semi-bandit feedback}, that is, payoffs at unguarded targets are interdependent and cannot be individually attributed.

\textbf{Challenge 3: Non-Stationary Payoffs.} Unlike classical settings with fixed target values, payoffs in HEC evolve as elephants adapt their strategies over time. The value function of a boundary segment depends on the current behavior of the elephant, which changes in response to observed guard patterns.

\begin{table}[h]
\centering
\caption{Comparison with Classical Security Games}
\label{tab:comparison}
\small
\begin{tabular}{lcc}
\toprule
\textbf{Aspect} & \textbf{Classical GSG} & \textbf{This Work} \\
\midrule
Adversary model & Known/learnable & Unknown \\
Payoffs & Known, fixed & Unknown, non-stationary \\
Feedback & Full or bandit & Semi-bandit, confounded \\
Solution concept & Equilibrium & Regret minimization \\
\bottomrule
\end{tabular}
\end{table}

These challenges necessitate an online learning approach that: (i) provides guarantees without behavioral assumptions, (ii) handles confounded observations where attack entry points are unidentifiable, and (iii) adapts to non-stationary payoffs through continuous exploration.

\subsection{Formal Game Definition}

\textbf{Players and Strategies.} Let $[N] = \{1, \ldots, N\}$ denote the set of boundary segments (targets). The defender has $K$ resources to protect at most $K$ targets. A defender's pure strategy is a binary vector $\mathbf{v} \in \{0,1\}^N$ where $v_i = 1$ if target $i$ is protected, with $\|\mathbf{v}\|_1 \leq K$. The set of all feasible pure strategies is denoted $[V]$. A mixed strategy is a probability distribution over $[V]$.

The attacker can target multiple segments simultaneously. Let $Q$ denote the maximum number of targets attackable per round. The attacker's pure strategy is $\mathbf{a} \in \{0,1\}^N$ where $a_i = 1$ if target $i$ is attacked, with $\|\mathbf{a}\|_1 \leq Q$. The set of attacker pure strategies is $[A]$.

\textbf{Payoffs and Utility.} The defender's payoff at target $i$ depends on coverage and attack status. Let $U_i^c$ denote the payoff when target $i$ is covered and attacked (successful interception), and $U_i^u$ the payoff when uncovered and attacked (crop damage). We have $U_i^u \leq U_i^c$ since interception is preferable, with payoffs bounded to $[-0.5, 0.5]$ for normalization. \textit{Critically, these payoffs are unknown to the defender a priori and must be learned online.}

Given strategies $\mathbf{v}_t$ and $\mathbf{a}_t$ at round $t$, the defender's utility is:
\begin{equation} \label{eq:utility_vanilla}
u(\mathbf{v}_t, \mathbf{a}_t) = \sum_{i \in [N]} v_{t,i} a_{t,i} U_i^c + \sum_{i \in [N]} (1 - v_{t,i}) a_{t,i} U_i^u
\end{equation}
Defining $r_{t,i} = a_{t,i}[U_i^c - U_i^u] \in [0,1]$ and $C(\mathbf{a}_t) = \sum_{i \in [N]} a_{t,i} U_i^u$, utility simplifies to:
\begin{equation}
u(\mathbf{v}_t, \mathbf{a}_t) = \mathbf{v}_t \cdot \mathbf{r}_t + C(\mathbf{a}_t)
\end{equation}

\textbf{Game Protocol and Objective.} The game proceeds for $T$ rounds. At each round $t$: (1) the defender selects $\mathbf{v}_t \in [V]$ based on history $F_{t-1}$; (2) the attacker selects $\mathbf{a}_t \in [A]$; (3) the defender receives utility $u(\mathbf{v}_t, \mathbf{a}_t)$ and observes partial feedback. We assume $\mathbf{r}_t$ may depend on history $F_{t-1}$ but not on the current action $\mathbf{v}_t$.

The objective is to minimize cumulative regret:
\begin{equation}
R_T = \max_{\mathbf{v} \in [V]} \sum_{t=1}^{T} \mathbf{r}_t \cdot \mathbf{v} - \mathbb{E}\left[\sum_{t=1}^{T} \mathbf{r}_t \cdot \mathbf{v}_t\right]
\end{equation}
where the first term is the utility of the optimal fixed hindsight strategy. Regret benchmarks against a fixed strategy because the optimal adaptive policy is unattainable when the adversary can select any $\mathbf{a}_t$ arbitrarily.

\textbf{Feedback Model.} In each round $t$, the defender observes:

\textit{Guarded segments} ($v_{t,i} = 1$): attack indicator $\mathbb{I}[a_{t,i} = 1]$ and reward $r_{t,i}$ if attacked, and

\textit{Unguarded segments} ($v_{t,i} = 0$): only aggregate crop damage $L_t = \sum_{i: v_{t,i}=0} a_{t,i} |U_i^u|$, with unidentified individual entry points.

This constitutes \textit{confounded semi-bandit feedback}: per-arm rewards are observed only for selected arms, while unselected arms yield an aggregate signal with unattributable components. Standard semi-bandit algorithms assume independent reward observation per selected action; our setting violates this, necessitating a payoff estimation mechanism that distributes observed losses across candidate entry points (Section~\ref{sec:adaptive-payoff-learning}).

\section{Resource Deployment for Strategic Guarding}

\subsection{Baseline: Follow-the-Perturbed-Leader Framework}

We build on the Follow-the-Perturbed-Leader with Uniform Exploration (FPL-UE) algorithm \cite{Haifeng2016}, which addresses online combinatorial optimization problems with semi-bandit feedback and unknown adversaries. We first review the key components of FPL-UE before presenting our adaptations.

\subsubsection{FPL Algorithm}
The Follow-the-Perturbed-Leader (FPL) algorithm \cite{Neu2013} solves online combinatorial optimization under semi-bandit feedback. In each round $t$, the learner selects an action from a finite set based on reward estimates $\hat{r}_{t,i}$ for each target $i$, initialized without prior knowledge ($\hat{r}_{1,i} = 0$). The defender's pure strategy $v_t$ is calculated as:
\begin{equation}
v_t = \arg\max_{v \in V} \{v \cdot (\hat{\mathbf{r}}_t + \mathbf{z})\},
\end{equation}
where $\mathbf{z}$ is a random perturbation vector with $z_i \sim \exp(\eta)$ drawn independently from an exponential distribution with parameter $\eta$. The reward estimate $\hat{r}_{t+1,i}$ depends on $p_{t,i}$, the probability of selecting the target $i$. Since $p_{t,i}$ cannot be computed efficiently in closed form, the Geometric Resampling (GR) algorithm with truncation parameter $M$ estimates $p_{t,i}$ by measuring reoccurrence times, effectively estimating $\frac{1}{p_{t,i}}$ as $P_{(t,i)}$.

\subsubsection{FPL-UE: Uniform Exploration Component}
FPL-UE \cite{Haifeng2016} extends FPL by introducing exploration to handle scenarios where $p_{t,i}$ estimates can be arbitrarily small. In each round $t$, the algorithm chooses between exploration and exploitation with probability $\gamma$. In the \textit{exploration phase}, a strategy is randomly selected from $\mathcal{E}_{\text{expl}} = \{v_1, \ldots, v_N\}$, where the target $i$ is protected in pure strategy $v_i$. Thus, each target is covered with probability $\frac{\gamma}{N}$. In the \textit{exploitation phase}, the optimized strategy $v_t$ is chosen based on reward estimate $\hat{\mathbf{r}}_t$ and random perturbation $\mathbf{z}$. The GR and FPL-UE algorithms are given in the Appendix~\ref{sec:gr-fpl-algos}.

\subsection{Adaptive Payoff Learning}
\label{sec:adaptive-payoff-learning}
A critical challenge in the HEC setting (Sect.~\ref{sec:challenge}) is that all the payoffs at the boundary patch are not independently observable. Although rewards $U^c_i$ from successful interceptions at covered targets can be directly observed, penalties $U^u_i$ at uncovered targets present a fundamental inference problem. When elephants successfully raid crops, the defender observes the damage of the agricultural plot but cannot determine which boundary patch was used for entry due to limited surveillance. Since elephants may access the same agricultural plot through multiple uncovered boundary patches, the payoff values are inherently interdependent, violating the independence assumption of classical security games.

\textbf{Dynamic Payoff Estimation Mechanism} To address payoff uncertainty, we develop a dynamic learning mechanism that updates reward and penalty estimates in real-time as the game progresses.

We assume that target payoffs follow an unknown distribution, with the defender receiving sample observations during gameplay. Let \texttt{MaxInterceptions} denote the maximum number of elephant trajectories intercepted at any protected target throughout all rounds, and let \texttt{MaxCropRaidLoss} denote the maximum crop damage observed in any round.

For target $i$ in round $t$, the covering reward $U^c_i$ is estimated from the number of intercepted trajectories, normalized by \texttt{MaxInterceptions} to yield $U^c_i \in [0, 0.5]$. The uncovering penalty $U^u_i$ is estimated from observed crop damage in round $t$, normalized by \texttt{MaxCropRaidLoss} to yield $U^u_i \in [-0.5, 0]$. Since we cannot identify which specific uncovered boundary patch was used, the total loss of crop raid is evenly distributed among all uncovered targets in the round $t$. This conservative approach ensures that all potentially vulnerable targets contribute to the penalty estimate, accommodating the partial observability constraint. The reward $r_{t,i} = a_{t,i}[U^c_i - U^u_i] \in [0,1]$ is then computed for each target and used to update the reward estimates $\hat{\mathbf{r}}_t$ in the online learning algorithm.

\subsection{The HERDS Algorithm}

\subsubsection{Adaptive Exploration-Exploitation Budget Allocation}
We modify the classical FPL-UE framework to accommodate the unique characteristics of HEC mitigation. Rather than allocating all $K$ resources to only exploration or exploitation in each round, HERDS dynamically partitions resources into exploration budget $K_{\text{expl}}$ and exploitation budget $K_{\text{expt}}$, where $K_{\text{expl}} + K_{\text{expt}} = K$ and $K_{\text{expl}}, K_{\text{expt}} \in \mathbb{Z}^+$. This resource partitioning is governed by an adaptive trade-off parameter $\gamma_{t+1}$ based on the observed crop damage and is defined as
\begin{equation} \label{eqn:gamma_adaptive}
\gamma_{t} = \frac{\texttt{CropRaidLoss}_{t-1}}{\texttt{MaxCropRaidLoss}},
\end{equation}
where the numerator is the crop damage in round $t-1$ and the denominator is the maximum damage observed before round $t$.

The exploration budget is set as $K_{\text{expl}} = \lfloor \gamma_t K \rfloor$ and the exploitation budget is set as $K_{\text{expt}} = K - K_{\text{expl}}$.

\textbf{Rationale:} Higher observed crop damage indicates ineffective strategies, which requires more exploration to identify optimal protection targets. On the other hand, lower damage suggests successful strategies, encouraging the exploitation of known effective targets. Unique to our HEC setting, this adaptive mechanism uses the observation of crop damage to dynamically balance exploration and exploitation.

\subsubsection{Target Selection Strategy}
In each round $t$, the first step involves calculating exploitation targets $[N_{\text{expt}}]$. This is done by considering $[V_{\text{expt}}]$, which comprises all pure strategies $v$ such that $\|v\|_1 = K_{\text{expt}}$. Then, the perturbation $\mathbf{z}$ is sampled with $z_i \sim \exp(\eta)$ for each $i$ in $[N]$. The optimal exploitation strategy, akin to FPL, is determined as follows
$
v_{\text{expt}} = \arg\max_{v \in V_{\text{expt}}} \{v \cdot (\hat{\mathbf{r}}_t + \mathbf{z})\}
$ which determines $[N_{\text{expt}}]$. Exploration target set $[N_{\text{expl}}]$ is formed by randomly sampling $K_{\text{expl}}$ targets from $[N] \setminus [N_{\text{expt}}]$. Finally, guards are deployed at all the selected exploration and exploitation targets.

\subsubsection{Complete HERDS Algorithm}
The Algorithm~\ref{algo:herds} presents the complete HERDS procedure, which integrates adaptive budget allocation, dynamic payoff estimation, and the modified FPL-UE framework. The key innovations in HERDS are adaptive resource partitioning based on observed crop damage (Lines 4, 14), 
simultaneous exploration and exploitation rather than probabilistic switching between modes (Lines 7--9), and dynamic payoff estimation that accounts for partial observability and interdependent target values (Lines 12--13). These modifications ensure that HERDS can effectively learn optimal guard deployment strategies against adaptive ecological adversaries with unknown behavior and uncertain, dynamically evolving payoffs. The standard Geometric Resampling (GR) algorithm (Algorithm~\ref{algo:gr} in Appendix~\ref{sec:gr-fpl-algos}) is used in Line 11. A visual representation of the algorithm for ease of understanding is given in Figure~\ref{fig:HERDS-block-diagram} of Appendix~\ref{herds-workflow}.

\begin{algorithm}
\caption{The HERDS Algorithm}
\label{algo:herds}
\begin{algorithmic}[1]

\REQUIRE $\eta \in \mathbb{R}^+, M \in \mathbb{Z}^+$;

\STATE Initialize the estimated reward $\hat{\mathbf{r}} = \mathbf{0} \in \mathbb{R}^N$;

\STATE Initialize $\gamma_{1} = 1$;

\FOR{$t = 1, \ldots, T$}

\STATE Calculate $K_{expl}=\lfloor\gamma_t K\rfloor$ and $K_{expt}=K-K_{expl}$;

\STATE Find $[V_{expt}]$, the set of all pure strategies of the defender such that $||\mathbf{v}||_1=K_{expt} \,\forall\, \mathbf{v}\in [V_{expt}]$;

\STATE Draw $z_i \sim \exp(\eta)$ independently for $i \in [N]$ and let $\mathbf{z} = (z_1, \ldots, z_{N})$;

\STATE Compute $v_{expt} = \arg\max_{v \in V_{expt}}\{v \cdot (\hat{\mathbf{r}} + \mathbf{z})\}$ to find $[N_{expt}]$;

\STATE Pick the exploration targets $[N_{expl}]$ by randomly sampling $K_{expl}$ targets from $[N] \setminus [N_{expt}]$; 

\STATE Form $v_t \in [0, 1]^N$ such that $v_t^{(i)}=1 \,, \forall i \in [N_{expt}]\cup[N_{expl}]$;

\STATE Adversary picks $a_t \in [0, 1]^N$ and defender plays $v_t$;

\STATE Run GR($\eta, M, \hat{\mathbf{r}}, t$): estimate $\frac{1}{p_{t,i}}$ as $P(t, i)$;

\STATE Update $\hat{\mathbf{r}}(t) \leftarrow \hat{\mathbf{r}}(t) + P(t, i) r_t(i) \cdot \mathbb{I}(t, i)$ where $\mathbb{I}(t, i) = 1$ for $i$ satisfying $v_{t,i} = 1$; $\mathbb{I}(t, i) = 0$ otherwise;

\STATE Calculate $\texttt{CropRaidLoss\_t}$ and Update $\texttt{MaxCropRaidLoss}$

\STATE Calculate $\gamma_{t+1} = \frac{\texttt{CropRaidLoss\_t}}{\texttt{MaxCropRaidLoss}}$

\ENDFOR

\end{algorithmic}
\end{algorithm}





\subsubsection{Performance Analysis}

HERDS inherits the theoretical foundation of FPL-UE \cite{Haifeng2016} while introducing adaptive exploration. We derive the following upper bound on cumulative regret.

\begin{equation} \label{eq:regret_v8}
\begin{split}
    R_T \leq 2T(1-\gamma_t)K\left(1-\frac{\gamma_t}{N}\right)^M + \frac{(1-\gamma_t)K (\log N + 1)}{\eta} + \\ \eta m T \min(m,(1-\gamma_t)K) + \sum_{t=1}^{T} \mathbb{E}\left[\gamma_t\right] K\,.
\end{split}
\end{equation}

The key departure from FPL-UE is our adaptive exploration parameter $\gamma_{t+1} = \frac{\texttt{CropRaidLoss\_t}}{\texttt{MaxCropRaidLoss}}$, which enables target-level sampling rather than strategy-level sampling. Although this adaptivity complicates the analysis, sublinear regret is guaranteed when $\gamma_t$ remains bounded by a constant $\hat{\gamma}$ (i.e., $\gamma_t \leq \hat{\gamma}$ for all $t$). The complete derivation is provided in the Appendix~\ref{derivation:regret-bound}.

Compared to FPL-UE, HERDS achieves two critical advantages through target-level sampling: (1) a higher minimum coverage probability per target, allowing reward estimates $\hat{r}$ to stabilize faster, and (2) accelerated decay of exploration-driven regret. As shown in Eqn.~\ref{eqn:HERDS-p(t,i)} (Appendix~\ref{derivation:regret-bound}), HERDS ensures that targets are never visited with a lower probability than in FPL-UE, leading to more comprehensive exploration and consequently to better reward estimates. This improved exploration reduces estimation bias, allowing HERDS to achieve comparable or better performance with smaller values of the truncation parameter $M$, which controls exploration depth in reward estimation. Our experiments confirm sublinear regret and demonstrate robustness across different values of $M$ (Figure~\ref{fig:regret-summary-plot}). An ablation study by modifying FPL-UE to use an adaptive $\gamma_t$ is given in Appendix~\ref{sec:ablation-studies}. We see that this addition improved FPL-UE, but still falls short of HERDS.

\section{Numerical Experiments}
\subsection{Experimental Setup}

\textbf{Simulation Environment.} Although field experiments would provide the most direct validation, the multi-stakeholder nature of HEC and the inherent risks to both humans and wildlife make direct field testing infeasible. Therefore, we employ a spatially-explicit prototype ABM \cite{Purathekandy2024} calibrated with elephant movement data from the Periyar-Agasthyamalai complex in India as a virtual laboratory for evaluation of HERDS. The ABM serves three critical purposes: (i) it provides a safe, repeatable testbed for policy learning without real-world consequences; (ii) it bypasses severe data constraints typical in wildlife domains by incorporating biologically-informed behavioral models; and (iii) it enables systematic evaluation against multiple adversarial strategies, which would be impossible to control in field settings.

\textbf{Game Configuration.} The continuous forest-agricultural boundary in the field of study (Periyar Agasthyamalai region of the Western Ghats in South India) is discretized into $N=57$ boundary segments. Each game proceeds for $T=100$ rounds, with each round simulating 10 days of elephant movement in the ABM. We vary the defender's resource budget $K \in \{3,4,5,6,7,8\}$ and the GR truncation parameter $M \in \{3,8,15\}$ to evaluate algorithm performance under different resource constraints and estimation accuracy requirements. The perturbation parameter is set to $\eta = 0.5$ across all experiments.

\textbf{Adversary Models.} Because HERDS is model-agnostic, we assess its robustness by evaluating it against two adversary models that span a range of rationality. These models are not assumptions embedded in our algorithm; instead, they serve as plausible representations of elephant behavior used to stress-test the learned policies.\\
\textbf{(i) Myopic Adversary Model (MAM):} Elephant agents have no memory or knowledge of guard patrol locations and treat all boundary segments indiscriminately. The adversary cannot observe the defender's strategy, representing the least strategic end of the rationality spectrum.\\
\textbf{(ii) Bounded Rationality Stackelberg Attacker Model (BRSAM):} Elephant agents initially lack knowledge of guard locations but build spatial memory through encounters during gameplay. Once they learn guard positions, they adaptively evade guarded segments and exploit unguarded areas. This model captures adaptive, memory-based behavior without perfect rationality, representing a more strategic adversary. Example trajectories for both models are shown in Appendix~\ref{app:mambrsam}. The complete implementation and source code for the MAM and BRSAM models are available in the supplementary materials. 


\textbf{Baseline Methods.} We compare HERDS with two baselines: FPL-UE \cite{Haifeng2016} and Static Policy. In the latter, a fixed guard deployment strategy generated from ABM simulations with partial observability, protecting the boundary segments most frequently attacked by elephants in historical data.

\textbf{Evaluation Metrics.} Performance is evaluated with three distinct metrics: \\
(i) \textbf{Cumulative Regret} ($R_T$) assesses the gap between the optimal hindsight strategy's utility and the algorithm's utility, indicating learning efficiency; \\
(ii) \textbf{Crop Raiding Loss} measures the kilograms of dry matter consumed or damaged by elephants penetrating defenses, reflecting real-world impact; \\
(iii) \textbf{Interception Rate} is the fraction of elephant movements diverted at guarded borders before entering farms, quantifying conflicts averted.



\subsection{Results}
Our experiments address four key research questions that validate our contributions.\\ 
    \textbf{RQ1:} \textit{Does HERDS achieve lower regret than baseline approaches, and how quickly does it converge?}\\
    \textbf{RQ2:} \textit{Does HERDS effectively reduce crop damage across different resource budgets?}\\
    \textbf{RQ3:} \textit{Does the adaptive payoff learning mechanism successfully identify high-value protection targets?}\\
    \textbf{RQ4:} \textit{How robust is HERDS against different adversary behavioral models?}
\subsubsection{RQ1: Learning Efficiency and Convergence}

\begin{figure}[h]
  \centering
  \includegraphics[width=\linewidth]{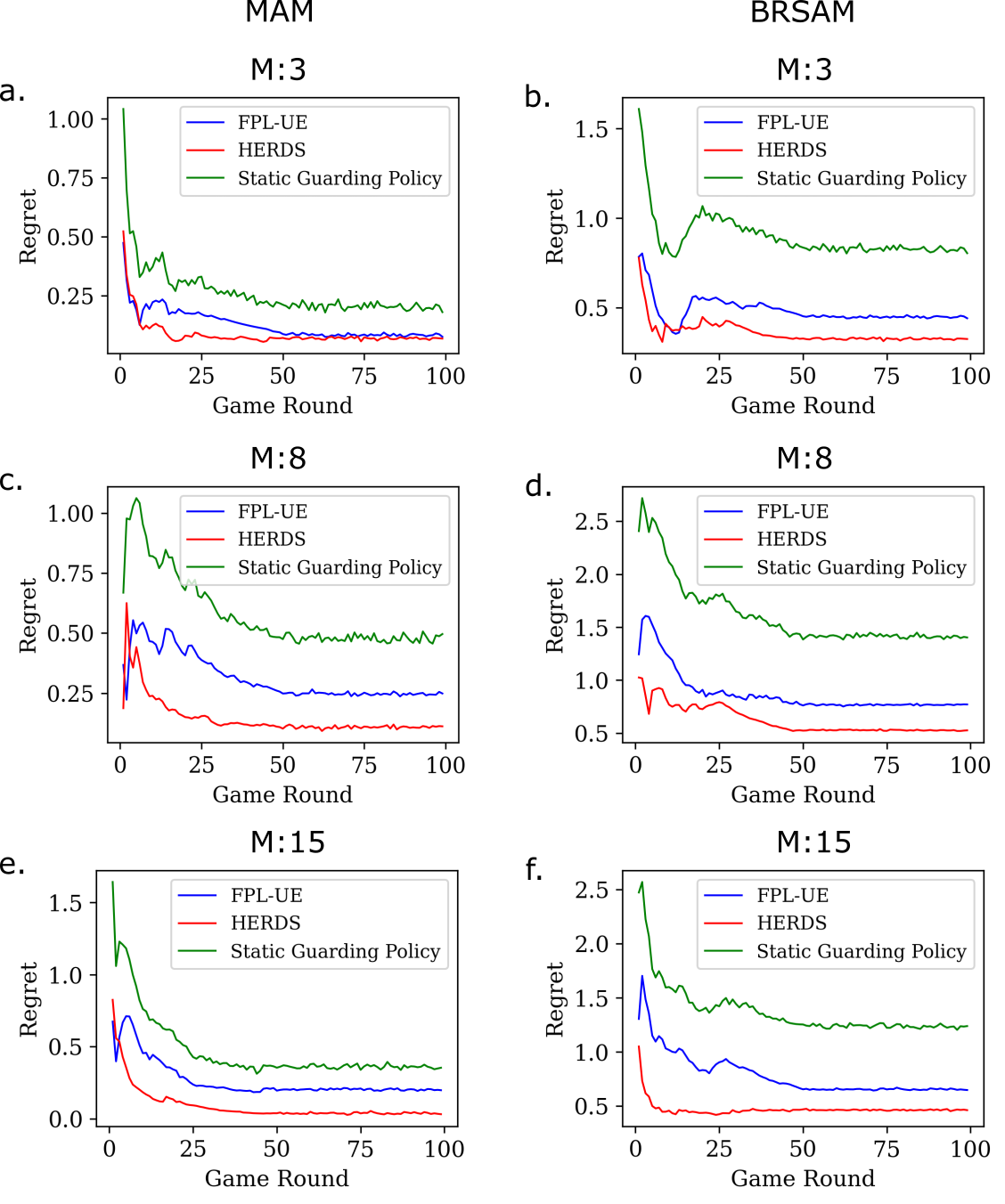}
  \caption{Cumulative regret comparison across 100 game rounds for HERDS, FPL-UE, and static guarding policy under different resource budgets and adversary models. Left column shows results against MAM (Myopic Adversary Model), right column against BRSAM (Bounded Rationality Stackelberg Attacker Model), with rows representing different GR truncation parameters. $K=5$ guards are used.}
  \label{fig:regret-summary-plot}
\end{figure}

Figure~\ref{fig:regret-summary-plot} presents the cumulative regret over 100 game rounds for HERDS, FPL-UE, and the static policy in different resource budgets and adversary models. We observe that HERDS consistently achieves lower regret than both baselines in all experimental configurations, with Table~\ref{tab:regret-summary-table} quantifying this advantage by showing that HERDS reduces final regret by 15-45\% compared to FPL-UE. Beyond overall performance, HERDS reaches near-optimal performance in approximately 40-50 rounds, while FPL-UE requires 60-80 rounds. Such a faster convergence is critical in real-world settings where early losses have significant socio-economic impacts. Finally, varying the truncation parameter $M$ (which affects probability estimation accuracy) shows that HERDS maintains stable performance even with less accurate estimates ($M=3$), while FPL-UE demonstrates degraded performance. The choice of $M$ implicitly controls the exploration depth in reward estimation, thereby affecting how the algorithm adapts to the adversary's behaviour. 

\begin{table*}
    \centering
    \caption{Regret values at the end of the game play against MAM and BRSAM adversary models. Lower is better.}
    \label{tab:regret-summary-table}
    \resizebox{2\columnwidth}{!}{
    \begin{tabular}{|c|*{12}{c|}} 
        \toprule
        \multicolumn{1}{|c|}{\textbf{K}} & 
        \multicolumn{4}{c|}{\textbf{M=3}} & \multicolumn{4}{c|}{\textbf{M=8}} & \multicolumn{4}{c|}{\textbf{M=15}} \\
        \midrule
        \multicolumn{1}{|c|}{} & 
        \multicolumn{2}{c|}{\textbf{MAM}} & \multicolumn{2}{c|}{\textbf{BRSAM}} & \multicolumn{2}{c|}{\textbf{MAM}} & \multicolumn{2}{c|}{\textbf{BRSAM}} & \multicolumn{2}{c|}{\textbf{MAM}} & \multicolumn{2}{c|}{\textbf{BRSAM}} \\
        \midrule
        \multicolumn{1}{|c|}{} & 
        \textbf{FPL-UE} & \textbf{HERDS} & \textbf{FPL-UE} & \textbf{HERDS} & \textbf{FPL-UE} & \textbf{HERDS} & \textbf{FPL-UE} & \textbf{HERDS} & \textbf{FPL-UE} & \textbf{HERDS} & \textbf{FPL-UE} & \textbf{HERDS} \\
        \midrule
        \textbf{3} & 0.101 & 0.077 & 0.222 & 0.097 & 0.177 & 0.079 & 0.321 & 0.214 & 0.157 & 0.051 & 0.387 & 0.314 \\
        \hline
        \textbf{4} & 0.111 & 0.083 & 0.341 & 0.153 & 0.248 & 0.099 & 0.411 & 0.342 & 0.186 & 0.055 & 0.511 & 0.428 \\
        \hline
        \textbf{5} & 0.118 & 0.101 & 0.487 & 0.245 & 0.268 & 0.116 & 0.671 & 0.528 & 0.213 & 0.078 & 0.673 & 0.518 \\
        \hline
        \textbf{6} & 0.243 & 0.122 & 0.499 & 0.269 & 0.311 & 0.132 & 0.677 & 0.574 & 0.342 & 0.087 & 0.741& 0.588 \\
        \hline
        \textbf{7} & 0.255 & 0.146 & 0.542 & 0.321 & 0.387 & 0.138 & 0.692 & 0.592 & 0.369 & 0.089 & 0.789 & 0.594 \\
        \hline
        \textbf{8} & 0.287 & 0.148 & 0.572 & 0.375 & 0.422 & 0.156 & 0.748 & 0.621 & 0.412 & 0.100 & 0.841 & 0.621 \\
        \bottomrule
    \end{tabular}
    }
\end{table*}

\begin{figure}[h]
  \centering
  \includegraphics[width=\linewidth]{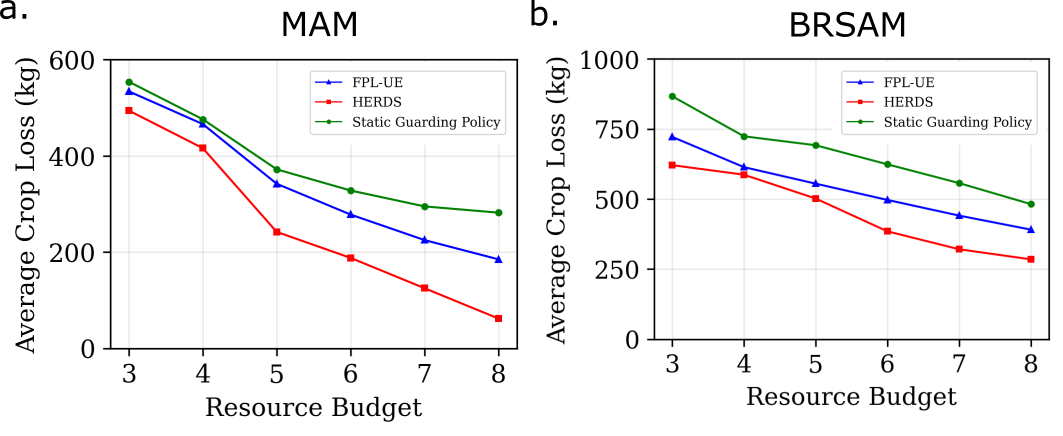}
  \caption{Average crop raiding loss (kg of dry matter) as a function of guard resource budget for (a) MAM and (b) BRSAM adversary models with $M=8$.}
  \label{fig:crop-raid-loss}
\end{figure}

\begin{figure}[h]
  \centering
  \includegraphics[width=0.925\linewidth]{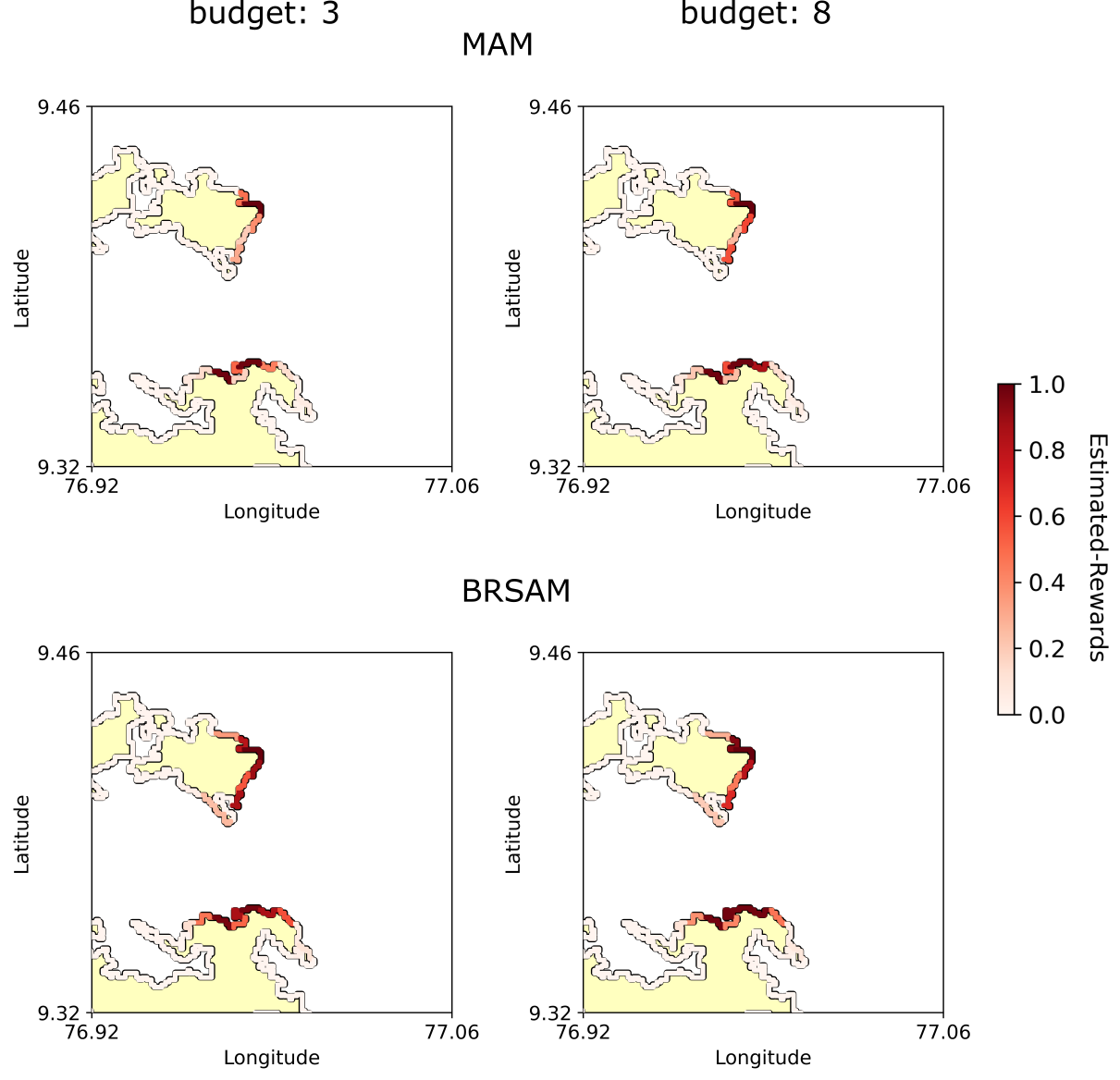}
  \caption{Spatial distribution of normalized estimated rewards for boundary segments at the end of 100-round gameplay, comparing scenarios with different resource budgets and adversary models (MAM top row, BRSAM bottom row). Color intensity represents reward magnitude, with warmer colors indicating higher strategic value.}
  \label{fig:estimated-rewards}
\end{figure}


\subsubsection{RQ2: Crop Damage Reduction}
While regret measures learning efficiency, crop damage quantifies real-world conservation impact. Figure \ref{fig:crop-raid-loss} shows the average crop raiding loss across different resource budgets for both adversary models, revealing that for the adaptive BRSAM adversary with $K=6$ guards, HERDS reduces average crop loss from 650 kg to 350 kg (46\% reduction) and is 30\% better than FPL-UE. HERDS outperforms baselines across all resource budgets. While all algorithms show diminishing returns as resources increase, HERDS reaches near-optimal damage reduction with fewer resources ($K=6$) compared to FPL-UE ($K=8$), suggesting more efficient resource utilization. Against the more intelligent BRSAM adversary, all methods experience higher crop loss, but HERDS maintains a larger performance advantage (20-40\% reduction) compared to its advantage against MAM (15-25\% reduction), demonstrating the value of simultaneous exploration and exploitation when facing adaptive opponents.

\subsubsection{RQ3: Adaptive Payoff Learning}
Figure \ref{fig:estimated-rewards} visualizes the normalized estimated rewards $\hat{r}_i$ for each boundary segment at the end of gameplay, revealing the spatial patterns learned by HERDS. The learned reward estimates are highly heterogeneous across boundary segments, with normalized values ranging from 0.25 to 1.0, validating our assumption that not all boundary patches have equal strategic importance. High-reward segments form distinct spatial clusters corresponding to proximity to water sources within agricultural areas, shorter distance to high-value crops, and terrain features facilitating elephant movement, with these patterns aligning with known ecological drivers of crop raiding \cite{Purathekandy2024,Shaffer2019}. The reward patterns differ between MAM and BRSAM scenarios, with high-reward regions shifting under BRSAM as elephants learn to avoid heavily guarded areas, demonstrating that HERDS successfully adapts its payoff estimates to changing adversary behavior. Our payoff learning mechanism (Section~\ref{sec:adaptive-payoff-learning}) successfully handles the partial observability constraint where specific entry points cannot be directly observed, building accurate reward estimates despite incomplete information by distributing penalties across uncovered segments.

\subsubsection{RQ4: Robustness to Adversary Models}
Analyzing performance against MAM and BRSAM adversaries sheds light on the robustness of the algorithm. HERDS outperforms baseline methods in both models, highlighting its resilience to attacker behavioral uncertainties. Against the tougher BRSAM adversary, the advantage of HERDS over FPL-UE enhances, indicating the value of simultaneous exploration-exploitation strategies against adaptive opponents. Figure~\ref{fig:crop-raid-loss} shows that static policies perform poorly against BRSAM (500-800 kg crop loss), as adaptive elephants quickly learn to bypass fixed guard locations, while HERDS stays effective (250-600 kg crop loss) by constantly updating deployment strategies.


\section{Conclusion}
This paper introduces the first online game-theoretic learning framework for strategic defense against adaptive ecological adversaries with unknown behavior and limited observability. We propose HERDS, which extends FPL-UE with dynamic exploration–exploitation partitioning, adaptive payoff estimation under confounded feedback, and joint resource deployment. The algorithm attains sublinear regret without relying on behavioral assumptions, a key property when adversary models are unavailable.

Experimental validation using an ABM calibrated with elephant movement data establishes three key findings: (1) Loss-driven budget allocation reduces regret by 15–45\% compared to uniform exploration, demonstrating that observed damage informs exploration strategy. (2) Distributing unobserved losses across candidate entry points successfully learns spatial patterns despite confounded semi-bandit feedback, with learned clusters aligning with ecological drivers. (3) Robust performance across adversary rationality levels achieves 40–50\% crop damage reduction and 30–50\% faster convergence (40–50 rounds versus 60–80).

Beyond HEC, our framework generalizes to security domains with strategic but unmodeled adversaries: anti-poaching patrols, infrastructure protection, and contested resource allocation. For conservation, HERDS provides actionable deployment strategies that require only partial observability. Future work should validate field performance, and incorporate guard effectiveness uncertainty. 

\section*{Impact Statement}

This work analyzes Human–Elephant Conflict using perspectives from security game theory and online machine learning, and, in doing so, proposes an algorithm that contributes to the broader field of machine learning. While our framework has the potential to support wildlife conservation and security efforts, it could also be misused for intrusive wildlife surveillance or coercive control systems. As the primary goal of conflict mitigation is to reduce harm to both local communities and vulnerable elephant populations, portraying wildlife as “adversaries” can be problematic, as it risks oversimplifying complex socio-ecological relationships. Any real-world deployment must involve the consent of affected communities and careful attention to animal welfare through extensive stakeholder dialogues. 





\bibliography{refs}

@inproceedings{Haifeng2016,
author = {Xu, Haifeng and Tran-Thanh, Long and Jennings, Nicholas R.},
title = {{Playing Repeated Security Games with No Prior Knowledge}},
year = {2016},
isbn = {9781450342391},
publisher = {International Foundation for Autonomous Agents and Multiagent Systems},
booktitle = {Proceedings of the 2016 International Conference on Autonomous Agents \& Multiagent Systems},
pages = {104–112},
numpages = {9},
location = {Singapore, Singapore},
series = {AAMAS '16}
}

@inproceedings{Shahrzad2019,
author = {Gholami, Shahrzad and Yadav, Amulya and Tran-Thanh, Long and Dilkina, Bistra and Tambe, Milind},
title = {{Don't Put All Your Strategies in One Basket: Playing Green Security Games with Imperfect Prior Knowledge}},
year = {2019},
isbn = {9781450363099},
publisher = {International Foundation for Autonomous Agents and Multiagent Systems},
booktitle = {Proceedings of the 18th International Conference on Autonomous Agents and MultiAgent Systems},
pages = {395–403},
numpages = {9},
location = {Montreal QC, Canada},
series = {AAMAS '19}
}

@book{Tambe2011, 
place={Cambridge}, 
title={{Security and Game Theory: Algorithms, Deployed Systems, Lessons Learned}}, publisher={Cambridge University Press}, 
author={Tambe, Milind}, 
year={2011}}

@article{Fang2016,
author = {Fang, Fei and Nguyen, Thanh H.},
title = {{Green security games: apply game theory to addressing green security challenges}},
year = {2016},
issue_date = {July 2016},
publisher = {Association for Computing Machinery},
address = {New York, NY, USA},
volume = {15},
number = {1},
doi = {10.1145/2994501.2994507},
journal = {SIGecom Exchanges},
pages = {78–83},
numpages = {6}
}

@article{An2011,
author = {An, Bo and Pita, James and Shieh, Eric and Tambe, Milind and Kiekintveld, Chris and Marecki, Janusz},
title = {{GUARDS and PROTECT: next generation applications of security games}},
year = {2011},
issue_date = {March 2011},
publisher = {Association for Computing Machinery},
address = {New York, NY, USA},
volume = {10},
number = {1},
doi = {10.1145/1978721.1978729},
journal = {SIGecom Exch.},
month = mar,
pages = {31–34},
numpages = {4}
}

@inproceedings{Fang2015,
author = {Fang, Fei and Stone, Peter and Tambe, Milind},
title = {{When security games go green: designing defender strategies to prevent poaching and illegal fishing}},
year = {2015},
isbn = {9781577357384},
publisher = {AAAI Press},
booktitle = {Proceedings of the 24th International Conference on Artificial Intelligence},
pages = {2589–2595},
numpages = {7},
location = {Buenos Aires, Argentina},
series = {IJCAI'15}
}

@inproceedings{Marecki2012,
title = {{Playing Repeated Stackelberg Games with Unknown Opponents}},
publisher = {International Foundation for Autonomous Agents and Multiagent Systems},
booktitle = {Proceedings of the 2014 International Conference on Autonomous Agents and Multi-Agent Systems},
author = {Marecki, Janusz and Tesauro, Gerry and Segal, Richard},
year = {2012},
series = {AAMAS '12},
pages = {821--828},
doi = {10.5555/2343776.2343814},
isbn = {978-0-9817381-2-3},
}

@inproceedings{Yang2014,
author = {Yang, Rong and Ford, Benjamin and Tambe, Milind and Lemieux, Andrew},
title = {{Adaptive resource allocation for wildlife protection against illegal poachers}},
year = {2014},
isbn = {9781450327381},
publisher = {International Foundation for Autonomous Agents and Multiagent Systems},
booktitle = {Proceedings of the 2014 International Conference on Autonomous Agents and Multi-Agent Systems},
pages = {453–460},
numpages = {8},
location = {Paris, France},
series = {AAMAS '14}
}

@InProceedings{Richard2014,
author="Kl{\'i}ma, Richard
and Kiekintveld, Christopher
and Lis{\'y}, Viliam",
editor="Poovendran, Radha
and Saad, Walid",
title="Online Learning Methods for Border Patrol Resource Allocation",
booktitle="Decision and Game Theory for Security",
year="2014",
publisher="Springer International Publishing",
address="Cham",
pages="340--349",
isbn="978-3-319-12601-2"
}

@inproceedings{Kar2015,
author = {Kar, Debarun and Fang, Fei and Delle Fave, Francesco and Sintov, Nicole and Tambe, Milind},
title = {{A Game of Thrones: When Human Behavior Models Compete in Repeated Stackelberg Security Games}},
year = {2015},
isbn = {9781450334136},
publisher = {International Foundation for Autonomous Agents and Multiagent Systems},
booktitle = {Proceedings of the 2015 International Conference on Autonomous Agents and Multiagent Systems},
pages = {1381–1390},
numpages = {10},
location = {Istanbul, Turkey},
series = {AAMAS '15}
}

@ARTICLE{Zhu2023,
author={Zhu, Jin and Zhang, Jinglong and Ling, Qiang and Dullerud, Geir E.},
journal={IEEE Transactions on Cognitive and Developmental Systems}, 
title={{Low Resource-Reallocation Defense Strategies for Repeated Security Games With No Prior Knowledge and Limited Observability}}, 
year={2023},
volume={15},
number={4},
pages={2156-2168},
doi={10.1109/TCDS.2023.3241364}}

@article{Xu2021, 
title={{Dual-Mandate Patrols: Multi-Armed Bandits for Green Security}}, 
volume={35}, 
DOI={10.1609/aaai.v35i17.17757}, 
number={17}, 
journal={Proceedings of the AAAI Conference on Artificial Intelligence}, 
author={Xu, Lily and Bondi, Elizabeth and Fang, Fei and Perrault, Andrew and Wang, Kai and Tambe, Milind}, 
year={2021}, 
month={May}, 
pages={14974-14982} }

@inproceedings{Qian2014,
author = {Qian, Yundi and Haskell, William B. and Jiang, Albert Xin and Tambe, Milind},
title = {{Online planning for optimal protector strategies in resource conservation games}},
year = {2014},
isbn = {9781450327381},
publisher = {International Foundation for Autonomous Agents and Multiagent Systems},
address = {Richland, SC},
booktitle = {Proceedings of the 2014 International Conference on Autonomous Agents and Multi-Agent Systems},
pages = {733–740},
numpages = {8},
location = {Paris, France},
series = {AAMAS '14}
}

@inproceedings{Paruchuri2007,
author = {Paruchuri, Praveen and Pearce, Jonathan P. and Tambe, Milind and Ordonez, Fernando and Kraus, Sarit},
title = {{An efficient heuristic approach for security against multiple adversaries}},
year = {2007},
isbn = {9788190426275},
publisher = {Association for Computing Machinery},
doi = {10.1145/1329125.1329344},
booktitle = {Proceedings of the 6th International Joint Conference on Autonomous Agents and Multiagent Systems},
articleno = {181},
numpages = {8},
location = {Honolulu, Hawaii},
series = {AAMAS '07}
}

@article{Fernando2008,
title = {{Review of Human-Elephant Conflict Mitigation Measures Practiced in South Asia}},
author = {Fernando, Prithiviraj and Kumar, M Ananda and Williams, A Christy and Wikramanayake, Eric and Aziz, Tariq and Singh, Sameer M},
journal = {WWF - World Wide Fund for Nature},
year={2008}, 
}

@article{Frank2010,
doi = {10.1371/journal.pone.0010688},
author = {Frank, David M. AND Sarkar, Sahotra},
journal = {PLOS ONE},
publisher = {Public Library of Science},
title = {{Group Decisions in Biodiversity Conservation: Implications from Game Theory}},
year = {2010},
month = {05},
volume = {5},
pages = {1-10},
number = {5},
}

@ARTICLE{Shaffer2019,
AUTHOR={Shaffer, L. Jen  and Khadka, Kapil K.  and Van Den Hoek, Jamon  and Naithani, Kusum J. }, 
TITLE={{Human-Elephant Conflict: A Review of Current Management Strategies and Future Directions}},
JOURNAL={Frontiers in Ecology and Evolution},
VOLUME={Volume 6 - 2018},
YEAR={2019},
DOI={10.3389/fevo.2018.00235},
ISSN={2296-701X},
}

@Article{Janssen2020,
AUTHOR = {Janssen, Stef and Matias, Diogo and Sharpanskykh, Alexei},
TITLE = {{An Agent-Based Empirical Game Theory Approach for Airport Security Patrols}},
JOURNAL = {Aerospace},
VOLUME = {7},
YEAR = {2020},
NUMBER = {1},
ARTICLE-NUMBER = {8},
ISSN = {2226-4310},
DOI = {10.3390/aerospace7010008}
}

@article{Adam2011,
title = {{The role of agent-based models in wildlife ecology and management}},
journal = {Ecological Modelling},
volume = {222},
number = {8},
pages = {1544-1556},
year = {2011},
issn = {0304-3800},
doi = {https://doi.org/10.1016/j.ecolmodel.2011.01.020},
author = {Adam J. McLane and Christina Semeniuk and Gregory J. McDermid and Danielle J. Marceau},
}

@inproceedings{Neu2013,
title = {{An Efficient Algorithm for Learning with Semi-Bandit Feedback}},
booktitle = {Algorithmic Learning Theory},
author = {Neu, Gergely and Bartók, Gábor},
editor = {Jain, Sanjay and Munos, Rémi and Stephan, Frank and Zeugmann, Thomas},
year = {2013},
pages = {234--248},
publisher = {Springer Berlin Heidelberg},
location = {Berlin, Heidelberg},
isbn = {978-3-642-40935-6}
}

@article{Purathekandy2024,
title = {{An agent-based model of elephant crop raid dynamics in the Periyar–Agasthyamalai complex, India}},
journal = {Ecological Modelling},
volume = {496},
pages = {110843},
year = {2024},
issn = {0304-3800},
doi = {https://doi.org/10.1016/j.ecolmodel.2024.110843},
author = {Anjali Purathekandy and Meera Anna Oommen and Martin Wikelski and Deepak N. Subramani},
}

@InProceedings{Kirkland2020,
author="Kirkland, Lisa-Ann
and de Waal, Alta
and de Villiers, Johan Pieter",
editor="Gerber, Aurona",
title="Evaluation of a Pure-Strategy Stackelberg Game for Wildlife Security in a Geospatial Framework",
booktitle="Artificial Intelligence Research",
year="2020",
publisher="Springer International Publishing",
address="Cham",
pages="101--118",
isbn="978-3-030-66151-9"
}

@article{Kalai2005,
  title={{Efficient algorithms for online decision problems}},
  author={Kalai, Adam Tauman and Vempala, Santosh S.},
  journal={Journal of Computer and System Sciences},
  year={2005},
  volume={71},
  number={3},
  pages={291--307},
  doi={10.1016/j.jcss.2004.10.016}
}

@book{CesaBianchi2006,
  title={{Prediction, Learning, and Games}},
  author={Cesa-Bianchi, Nicol{\`o} and Lugosi, G{\'a}bor},
  year={2006},
  publisher={Cambridge University Press},
  address={Cambridge},
  isbn={978-0-521-84108-5}
}

@article{Audibert2010,
  author  = {Audibert, Jean-Yves and Bubeck, S{\'e}bastien},
  title   = {{Regret Bounds and Minimax Policies under Partial Monitoring}},
  journal = {Journal of Machine Learning Research},
  year    = {2010},
  volume  = {11},
  pages   = {2785--2836}
}

@article{Auer2002,
  author = {Auer, Peter and Cesa-Bianchi, Nicol{\`o} and Fischer, Paul},
  title = {{Finite-time Analysis of the Multiarmed Bandit Problem}},
  journal = {Machine Learning},
  year = {2002},
  volume = {47},
  number = {2--3},
  pages = {235--256},
  doi = {10.1023/A:1013689704352}
}

@article{Awerbuch2008,
  title={{Online linear optimization and adaptive routing}},
  author={Awerbuch, Baruch and Kleinberg, Robert D.},
  journal={Journal of Computer and System Sciences},
  year={2008},
  volume={74},
  number={1},
  pages={97--114},
  doi={10.1016/j.jcss.2007.04.016}
}

@inproceedings{McMahan2004,
  author    = {McMahan, H. Brendan and Blum, Avrim},
  title     = {{Online Geometric Optimization in the Bandit Setting Against an Adaptive Adversary}},
  booktitle = {Proceedings of the 17th Annual Conference on Learning Theory (COLT)},
  year      = {2004},
  pages     = {109--123},
  publisher = {Springer},
  series    = {Lecture Notes in Computer Science},
  volume    = {3120},
  doi       = {10.1007/978-3-540-27819-1_8}
}
\bibliographystyle{icml2026}

\newpage
\appendix
\onecolumn

\section{Symbols and Notations}

Table~\ref{tab:notation} contains a summary of the symbols and notations used in the paper.

\begin{table}[h!]
\centering
\caption{Notation Summary}
\label{tab:notation}
\resizebox{0.75\columnwidth}{!}{
\begin{tabular}{cl}
\hline
\textbf{Notation} & \textbf{Description} \\
\hline
\multicolumn{2}{l}{\textit{Game Setup}} \\
$N$ & Number of boundary segments (targets) \\
$K$ & Number of deployable guard resources \\
$T$ & Total number of game rounds \\
$Q$ & Maximum number of boundary segments elephants can target per round \\
$\Delta L$ & Fixed length of each boundary segment \\
$[N]$ & Set of all targets $\{1, 2, \ldots, N\}$ \\
\hline
\multicolumn{2}{l}{\textit{Strategies}} \\
$\mathbf{v} \in \{0,1\}^N$ & Defender's pure strategy (binary vector) \\
$V$ & Set of all pure strategies of the defender \\
$\mathbf{a} \in \{0,1\}^N$ & Attacker's pure strategy (binary vector) \\
$A$ & Set of all pure strategies of the attacker \\
$\mathbf{v}_t$ & Defender's strategy at round $t$ \\
$\mathbf{a}_t$ & Attacker's strategy at round $t$ \\
$v_{t,i}$ & Binary indicator: 1 if target $i$ is protected at round $t$, 0 otherwise \\
$a_{t,i}$ & Binary indicator: 1 if target $i$ is attacked at round $t$, 0 otherwise \\
\hline
\multicolumn{2}{l}{\textit{Payoffs and Utilities}} \\
$U^c_i$ & Payoff for covering (protecting) target $i$ \\
$U^u_i$ & Payoff for not covering target $i$ \\
$u(\mathbf{v}_t, \mathbf{a}_t)$ & Utility of the defender at round $t$ \\
$r_{t,i}$ & $a_{t,i}[U^c_i - U^u_i]$ \\
$\hat{r}_{t,i}$ & Estimated reward for target $i$ at round $t$ \\
$\hat{\mathbf{r}}_t$ & Estimated reward vector at round $t$ \\
$C(\mathbf{a}_t)$ & Constant term in utility function, $C(\mathbf{a}_t) = \sum_{i \in [N]} a_{t,i} U^u_i$ \\
\hline
\multicolumn{2}{l}{\textit{Game History and Learning}} \\
$F_t$ & Complete history of the game up to and including round $t$ \\
$F_0$ & No initial history information \\
$R_T$ & Regret estimate at round $T$ \\
\hline
\multicolumn{2}{l}{\textit{Algorithm Parameters}} \\
$\eta$ & Perturbation parameter for exponential distribution \\
$\mathbf{z}$ & Random perturbation vector \\
$z_i$ & Perturbation for target $i$, where $z_i \sim \exp(\eta)$ \\
$M$ & Truncation parameter for Geometric Resampling (GR) algorithm \\
$P(t,i)$ & Estimate of $\frac{1}{p_{t,i}}$ from GR algorithm \\
$p_{t,i}$ & Probability of selecting target $i$ at round $t$ \\
$\gamma_t$ & Exploration-exploitation trade-off parameter at round $t$ \\
$K_{\text{expl}}$ & Exploration resource budget \\
$K_{\text{expt}}$ & Exploitation resource budget \\
$[N_{\text{expt}}]$ & Set of targets selected for exploitation \\
$[N_{\text{expl}}]$ & Set of targets selected for exploration \\
$[V_{\text{expt}}]$ & Set of all pure strategies with $\|\mathbf{v}\|_1 = K_{\text{expt}}$ \\
$E$ & Set of exploration strategies \\
$\mathbb{I}(t,i)$ & Indicator function: 1 if $v_{t,i} = 1$, 0 otherwise \\
\hline
\multicolumn{2}{l}{\textit{Performance Metrics}} \\
\texttt{MaxInterceptions} & Maximum number of elephant trajectories intercepted at any target \\
\texttt{MaxCropRaidLoss} & Maximum crop damage (kg) observed in any round \\
$\texttt{CropRaidLoss}_t$ & Crop raiding loss (kg of dry matter) at round $t$ \\
\hline
\end{tabular}
}
\end{table}
\section{Additional Figures}

\subsection{Schematic Representation of the HERDS Workflow} \label{herds-workflow}

\begin{figure}
  \centering
  \includegraphics[width=0.585\linewidth]{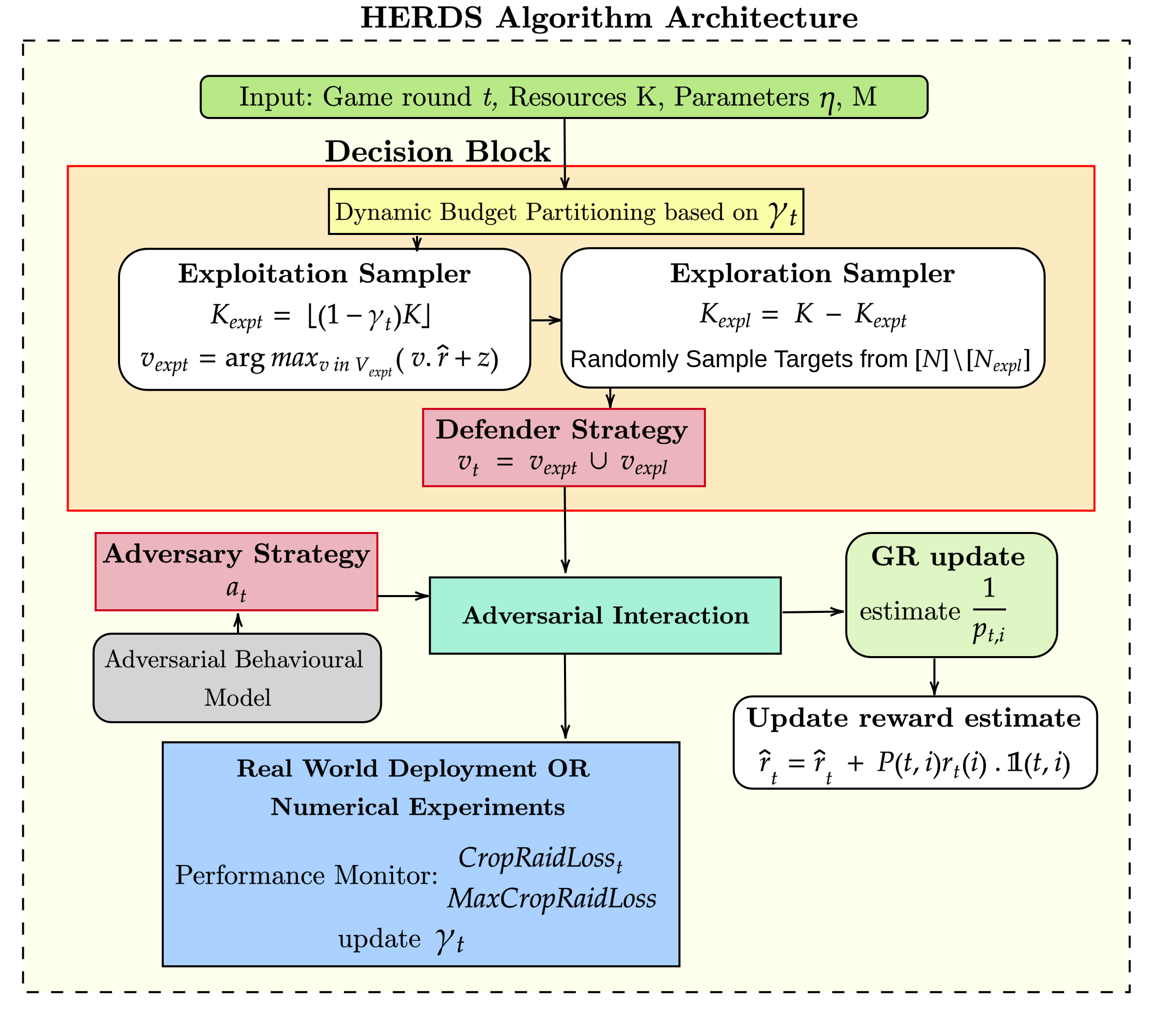}
  \caption{Flowchart with components of our HERDS algorithm. }
  \label{fig:HERDS-block-diagram}
\end{figure}

The components of the proposed HERDS algorithm are illustrated in Figure~\ref{fig:HERDS-block-diagram}.

\begin{figure}
  \centering
  \includegraphics[width=0.65\linewidth]{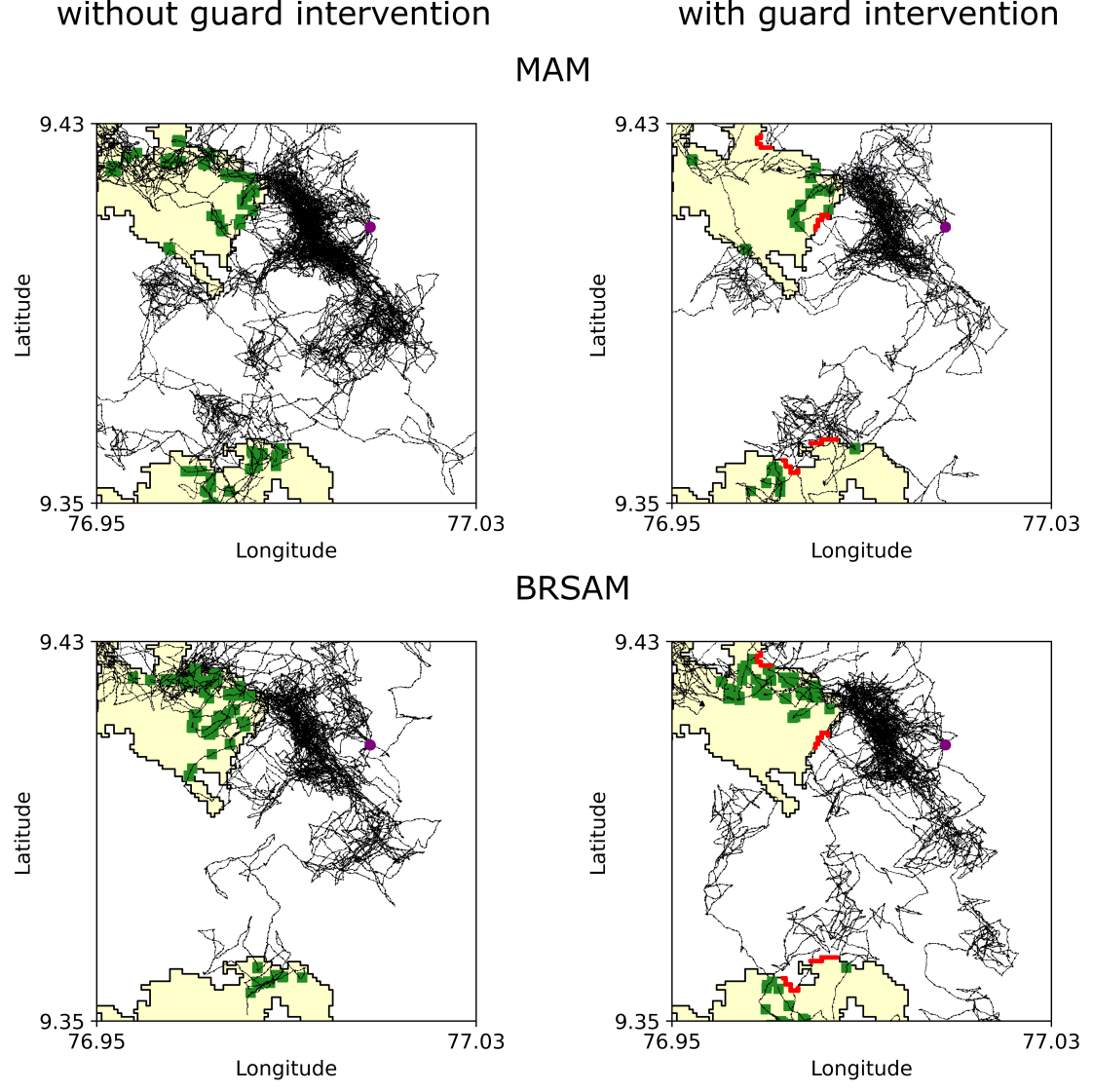}
  \caption{Agent-based simulations of elephant trajectories comparing no guard intervention (left) versus guard intervention (right) for two adversary models: Myopic Adversary Model (MAM, top) and Bounded Rationality Stackelberg Attacker Model (BRSAM, bottom). Purple circles mark elephant starting locations, black lines show trajectories, yellow regions are human settlements, white regions are forest, green squares are crop-raided agricultural plots, and red patches are guarded boundary segments. }
  \label{fig:trajectory-summary}
\end{figure}

\subsection{Trajectory Simulations of Attacker Models} \label{app:mambrsam}

Figure~\ref{fig:trajectory-summary} shows examples of trajectory evolution for the two attacker models, Myopic Adversary Model (MAM) and Bounded Rationality Stackelberg Attacker Model (BRSAM), used in this paper. Spatial patterns demonstrate how adaptive adversaries (BRSAM) alter their approach routes in response to guard deployments, motivating the need for dynamic defense strategies. These behavioral models ensure robustness and generalizability of the guard deployment model against a wide range of potential adversaries while accounting for uncertainty regarding actual elephant decision-making in the field.

\section{GR and FPL-UE Algorithms} \label{sec:gr-fpl-algos}


Algorithm~\ref{algo:gr} lists the GR algorithm and algorithm~\ref{algo:fplue} lists the FPL-UE algorithm. These are used in our HERDS algorithm listed in algorithm~\ref{algo:herds} given in the main paper.

\begin{algorithm}[h!]
\caption{The GR Algorithm}\label{algo:gr}
\begin{algorithmic}[1]
\REQUIRE $\eta \in \mathbb{R}^+, M \in \mathbb{Z}^+, \hat{\mathbf{r}} \in \mathbb{R}^n, t \in \mathbb{N}$;
\ENSURE $P(t) := \{P(t, 1), \ldots, P(t, n)\} \in \mathbb{Z}^N$
\STATE Initialize $\forall i \in [N]: P(t, i) = 0, p = 1$;
\FOR{$p = 1, \ldots, M$}
\STATE Repeat steps 4 to 9 using $\gamma_t$ in Algorithm 3 once to produce $\hat{v}$ as a simulation of $v_t$;
\FOR{all $i \in [N]$}
\IF{$p < M$ and $\hat{v}_i = 1$ and $P(t, i) = 0$}
\STATE $P(t, i) = p$;
\ELSIF{$p = M$ and $P(t, i) = 0$}
\STATE Set $P(t, i) = M$;
\ENDIF
\ENDFOR
\IF{$P(t, i) > 0$ for all $i \in [N]$} \STATE Break; \ENDIF
\ENDFOR
\end{algorithmic}
\end{algorithm}

\begin{algorithm}[h!]
\caption{The FPL-UE Algorithm}\label{algo:fplue}
\begin{algorithmic}[1]
\REQUIRE $\eta \in \mathbb{R}^+, M \in \mathbb{Z}^+, \gamma \in [0, 1]$;
\STATE Initialize the estimated reward $\hat{\mathbf{r}} = \mathbf{0} \in \mathbb{R}^N$;
\STATE Pick the set of exploration strategies $E = \{v_1, \ldots, v_n\}$ such that target $i$ is protected in pure strategy $v_i$;
\FOR{$t = 1, \ldots, T$}
\STATE Sample $flag \in [0, 1]$ such that $flag = 0$ with prob. $\gamma$;
\IF{$flag = 0$}
\STATE Let $v_t$ be a uniform randomly sampled strategy from $E$;
\ELSE
\STATE Draw $z_i \sim \exp(\eta)$ independently for $i \in [N]$ and let $\mathbf{z} = (z_1, \ldots, z_n)$;
\STATE $v_t = \arg\max_{v \in V}\{v \cdot (\hat{\mathbf{r}} + \mathbf{z})\}$;
\ENDIF
\STATE Adversary picks $a_t \in [0, 1]^N$ and defender plays $v_t$;
\STATE Run GR($\eta, M, \hat{\mathbf{r}}, t$): estimate $\frac{1}{p_{t,i}}$ as $P(t, i)$;
\STATE Update $\hat{\mathbf{r}}(t) \leftarrow \hat{\mathbf{r}}(t) + P(t, i) r_t(i) \cdot \mathbb{I}(t, i)$ where $\mathbb{I}(t, i) = 1$ for $i$ satisfying $v_{t,i} = 1$; $\mathbb{I}(t, i) = 0$ otherwise;
\ENDFOR
\end{algorithmic}
\end{algorithm}
\clearpage
\section{Regret Bound for HERDS}
\label{derivation:regret-bound}

\subsection{Preliminaries}

The defender's utility is at round $t$:

\begin{equation} \label{eq:utility_v1}
u(\mathbf{v}_t, \mathbf{a}_t) = \sum_{i \in [N]} v_{t,i} a_{t,i} U_i^c + \sum_{i \in [N]} (1 - v_{t,i}) a_{t,i} U_i^u
\end{equation}

Here $[N] = \{1, \ldots, N\}$ denote the set of boundary segments (targets), $\mathbf{v}_t$ is the defender strategy at $t$, $\mathbf{a}_t$ is the attacker strategy at $t$, $U_i^c$ denotes the payoff for covering target $i$, $U_i^u$ denotes the payoff for uncovering target $i$. We also assume $U_i^u \leq U_i^c$ since covering a target is always better than not covering it, with bounded payoffs ($U_i^c$, $U_i^u$ $\in$ $[-0.5, 0.5]$) for normalization.

Simplifying, 

\begin{equation} \label{eq:utility_v2}
u(\mathbf{v}_t, \mathbf{a}_t) = \sum_{i \in [N]} v_{t,i} a_{t,i} (U_i^c - U_i^u) + \sum_{i \in [N]}a_{t,i} U_i^u
\end{equation}

Defining $r_{t,i} = a_{t,i}[U_i^c - U_i^u] \in [0,1]$ and $C(\mathbf{a}_t) = \sum_{i \in [N]} a_{t,i} U_i^u$, utility simplifies to:
\begin{equation} \label{eq:utility_v3}
u(\mathbf{v}_t, \mathbf{a}_t) = \mathbf{v}_t \cdot \mathbf{r}_t + C(\mathbf{a}_t)
\end{equation}

Regret $R_T$ at step $T$ is defined as the difference between the total cumulative utility of the best fixed hindsight strategy and the expected cumulative utility of the algorithm used:

\begin{equation} \label{eq:regret_v1}
R_T = \max_{\mathbf{v} \in [V]} \sum_{t=1}^{T} u(\mathbf{v}, \mathbf{a}_t) - \mathbb{E}\left[\sum_{t=1}^{T} u(\mathbf{v}_t, \mathbf{a}_t)\right]
\end{equation}

\begin{equation} \label{eq:regret_v2}
R_T = \max_{\mathbf{v} \in [V]} \sum_{t=1}^{T} (\mathbf{v} \cdot \mathbf{r}_t + C(\mathbf{a}_t)) - \mathbb{E}\left[\sum_{t=1}^{T} (\mathbf{v}_t \cdot \mathbf{r}_t + C(\mathbf{a}_t))\right]
\end{equation}


\begin{equation} \label{eq:regret_v4}
R_T = \max_{\mathbf{v} \in [V]} \sum_{t=1}^{T} \mathbf{r}_t \cdot \mathbf{v} - \mathbb{E}\left[\sum_{t=1}^{T} \mathbf{r}_t \cdot \mathbf{v}_t\right]
\end{equation}

The optimal fixed hindsight strategy $v*$ is defined as
\begin{equation} \label{eq:optimal-hindsight-strategy}
v* = \arg \max_{\mathbf{v} \in [V]} \sum_{t=1}^{T} \mathbf{r}_t \cdot \mathbf{v} 
\end{equation}  

\subsection{Lemmas}

The following lemmas from Xu et al., \yrcite{Haifeng2016} are used:

\begin{itemize}

\item The GR algorithm produces estimates $\hat{r}_{t,i}$ such that

\begin{equation} \label{eq:lemma-bias-estimate}
\mathbb{E}[\hat{r}_{t,i} \mid \mathcal{F}_{t-1}] = \bigl(1 - (1 - p_{t,i})^{M}\bigr) r_{t,i}
\end{equation} 

where, $\mathcal{F}_{t-1}$ is the history up to time $t-1$, $p_{t,i} = \Pr[v_{t,i} = 1]$ is the probability target $i$ is protected at time $t$, $M$ is the truncation parameter in the GR algorithm, $r_{t,i}$ is the true reward at the target $i$ in round $t$, and $\hat{r}_{t,i}$ is the estimation of reward at the target $i$ in round $t$.

Here, 
\begin{equation} \label{eq:reward-estimate}
\hat{r}_{t,i} \leftarrow \hat{r}_{t,i} + P_{t,i} r_{t,i} \mathbb{I}(t, i) = 1 
\end{equation}

where $P_{t,i}$ is the estimate of $\frac{1}{p_{t,i}}$ and $\mathbb{I}(t, i)$ is the indicator function that is 1 if target $i$ is chosen by the defender at round $t$ and 0 otherwise.

In FPL-UE, the algorithm explores with probability $\gamma$ and exploits with probability $1 - \gamma$. Since each target is protected by at least one pure strategy in the set of exploration strategies, $p_{t,i}$ is lower bounded as

\begin{equation} \label{eqn:FPLUE-p(t,i)}
p_{t,i} \geq \frac{\gamma}{n} 
\end{equation}

In HERDS, which explores some targets while exploiting others within each round, the algorithm maintains a lower bound on the protection probability $p_{t,i}$ as

\begin{equation} \label{eqn:HERDS-p(t,i)}
p_{t,i} \geq \frac{\gamma_t}{N-K_{expt}} \geq \frac{\gamma_t}{N-(1-\gamma_t)K}
\end{equation}

where ${\gamma_t}$ is the exploration-exploitation trade-off parameter at round $t$, and $K$ is the total number of deployable guard resources.

This lower bound ensures that even targets that appear bad based on current history are occasionally sampled, which is necessary to maintain an unbiased reward estimate.

\item Also, for any round $t$, the reward vector $r_t$ satisfies 

\begin{equation} \label{eqn:lemma-r_t}
\|r_t\|_1 \leq m
\end{equation}

where, $r_{t,i} = a_{t,i}[U_i^c - U_i^u]$ with $U_i^c, U_i^u \in [-0.5, 0.5]$, $a_t \in \{0,1\}^N$ is the attacker's pure strategy at round $t$, and $m$ is the maximum number of targets the attacker can attack simultaneously ($\|a_t\|_1 \leq m$). 

In HERDS, $r_{t,i} \in [0,1]$. Therefore, $\|r_{t,i}\|_1 \leq N$. But we can assume that  $\|r_{t,i}\|_1 \leq m$ where $m << N$ because of the constraints on the attacker's movement. In the worst-case scenario, $m = N$.

\end{itemize}

\subsection{Regret Decomposition}

As in the previous literature \cite{Haifeng2016}, let us define some notation to analyze the HERDS algorithm. Let $\tilde{v_t}$ denote the strategy HERDS would choose if it only did exploitation with all $K_{expt}$ resources ($K_{expt} = \lfloor (1-\gamma_t) K \rfloor$) (i.e., $\tilde{v_t}$ = $v_{expt}$). 

Given the optimal fixed hindsight strategy $v*$ (Eqn.~\ref{eq:optimal-hindsight-strategy}), we can rewrite Eqn.~\ref{eq:regret_v4} as

\begin{equation} \label{eq:regret_v5}
R_T = \sum_{t=1}^{T} \mathbf{r}_t \cdot \mathbf{v*} - \mathbb{E}\left[\sum_{t=1}^{T} \mathbf{r}_t \cdot \mathbf{v}_t\right]
\end{equation}

Let us add and subtract $\sum_{t=1}^{T} \mathbf{r}_t \cdot\tilde{v_t}$ from Eqn.~\ref{eq:regret_v5}

\begin{equation} \label{eq:regret_v6}
R_T = \left( \sum_{t=1}^{T} \mathbf{r}_t \cdot \mathbf{v*} - \sum_{t=1}^{T} \mathbf{r}_t \cdot\tilde{v_t} \right) + \left( \sum_{t=1}^{T} \mathbf{r}_t \cdot\tilde{v_t} - \mathbb{E}\left[\sum_{t=1}^{T} \mathbf{r}_t \cdot \mathbf{v}_t\right]\right)
\end{equation}  

Thus, we can decompose the regret $R_t$ into two components:

\begin{equation} \label{eq:regret_v7}
R_T = R_T^{a} + R_T^{b}
\end{equation}  

Let us consider each individual component in the Eqn.~\ref{eq:regret_v7}:

\begin{itemize}
    \item[1.]
    \begin{equation} \label{eq:regret_exploitation_part_v1}
        R_T^{a} = \sum_{t=1}^{T} \mathbf{r}_t \cdot \mathbf{v*} - \sum_{t=1}^{T} \mathbf{r}_t \cdot\tilde{v_t} 
    \end{equation}

    Here, $K_{expt} = \lfloor (1-\gamma_t)K \rfloor$ resources are chosen using the FPL-based algorithm, with $K_{expt}$ varying with the time step $t$. 

    For the FPL algorithm, the upper bound for the regret from the exploitation part is given as follows \cite{Haifeng2016}:
    
    \begin{equation} \label{eq:regret_exploitation_part_v2}
        R_T^{FPL} \leq 2Tk(1-\frac{\gamma}{N})^M + \frac{k (\log N + 1)}{\eta} + \eta m T min(m,k)
    \end{equation}

    In HERDS, $k = K_{expt}$  and $K_{expt} \leq K$. Also, $\gamma = \gamma_t$.

    $$K_{expt} = \lfloor (1-\gamma_t) K \rfloor \approx (1-\gamma_t)K$$

    \begin{equation} \label{eq:regret_exploitation_part_v3}
        R_T^{a} \leq 2TK_{expt}(1-\frac{\gamma_t}{N})^M + \frac{K_{expt} (\log N + 1)}{\eta} + \eta m T min(m,K_{expt})
    \end{equation}

    \begin{equation} \label{eq:regret_exploitation_part_v4}
        R_T^{a} \leq 2T(1 - \gamma_t)K(1-\frac{\gamma_t}{N})^M + \frac{(1 - \gamma_t) K (\log N + 1)}{\eta} + \eta m T min(m,(1-\gamma_t)K)
    \end{equation}
    
    \item[2.] 
    \begin{equation} \label{eq:regret_exploration_part_v1}
        R_T^{b} = \sum_{t=1}^{T} \mathbf{r}_t \cdot\tilde{v_t} - \mathbb{E}\left[\sum_{t=1}^{T} \mathbf{r}_t \cdot \mathbf{v}_t\right] 
    \end{equation}

    Here, $K_{expl} = \lfloor \gamma_t K \rfloor$ resources are chosen randomly from N, not optimally. 

    Each exploration resource could yield up to 1 less reward ($r_{t,i} \in [0,1]$) than the exploitation resource it replaced. Therefore the loss in each round $\leq K_{expl}$

    \begin{equation} \label{eq:regret_exploration_part_v2}
        R_T^{b} \leq \mathbb{E}\left[\sum_{t=1}^{T} K_{expl}\right]
    \end{equation}

    \begin{equation} \label{eq:regret_exploration_part_v3}
        R_T^{b} \leq \mathbb{E}\left[\sum_{t=1}^{T} \gamma_tK\right] 
    \end{equation}

    \begin{equation} \label{eq:regret_exploration_part_v4}
        R_T^{b} \leq \sum_{t=1}^{T} \mathbb{E}\left[\gamma_t\right] K
    \end{equation}

\end{itemize}

\subsection{Total Regret Bound}

Putting all together, we have the following upper bound for the regret $R_T$ in HERDS:

    \begin{equation} 
        R_T \leq 2T(1-\gamma_t)K(1-\frac{\gamma_t}{N})^M + \frac{(1-\gamma_t)K (\log N + 1)}{\eta} + \eta m T min(m,(1-\gamma_t)K) + \sum_{t=1}^{T} \mathbb{E}\left[\gamma_t\right] K
    \end{equation}

The first three terms in Eqn.~\ref{eq:regret_v8} are very similar to the exploitation regret terms in Xu et al.~\yrcite{Haifeng2016}. This is expected as HERDS is a variant of FPL-UE with adaptive exploration. The adaptive part $\gamma_t$ adds complexity to the analysis, but in the worst-case analysis, if it can be bounded by a $\hat{\gamma}$ ($\gamma \leq \hat{\gamma}$), we can ensure that the HERDS algorithm has sublinear regret. 

By taking $\gamma_t \approx \mathcal{O}(\frac{1}{\sqrt(T)})$, we can obtain the upper bound for the regret for HERDS as $\mathcal{O}\sqrt(T)$.

\section{Ablation Studies} \label{sec:ablation-studies}

To empirically validate the structural integrity of our proposed framework and the contribution of components within the HERDS algorithm, we performed an ablation study by modifying the FPL-UE algorithm to use an adaptive $\gamma_t$ (Eqn.~\ref{eqn:gamma_adaptive}) based on observed crop raid losses (FPL-UE-A). The results of this ablation study are summarized in Table~\ref{tab:ablation_study}. Because of improved exploration, there is a faster reduction in the regret value in the FPL-UE-A when compared to the FPL-UE algorithm. However, in the FPL-UE-A, which still uses the strategy-level sampling, even with adaptive $\gamma_t$, the final regret values are similar to those in FPL-UE. This demonstrates that the advantage of HERDS lies in target-level rather than strategy-level sampling, with improved exploration leading to rapid regret minimization and target-level sampling resulting in comparatively lower regret. 

\begin{table*}[h!]
    \centering
    \caption{Ablation study with adaptive exploration-exploitation in FPL-UE (FPL-UE-A)}
    \label{tab:ablation_study}
    \resizebox{\columnwidth}{!}{
    \begin{tabular}{|c|*{12}{c|}} 
        \toprule
        \multicolumn{1}{|c|}{\textbf{K}} & 
        \multicolumn{4}{c|}{\textbf{M=3}} & \multicolumn{4}{c|}{\textbf{M=8}} & \multicolumn{4}{c|}{\textbf{M=15}} \\
        \midrule
        \multicolumn{1}{|c|}{} & 
        \multicolumn{2}{c|}{\textbf{MAM}} & \multicolumn{2}{c|}{\textbf{BRSAM}} & \multicolumn{2}{c|}{\textbf{MAM}} & \multicolumn{2}{c|}{\textbf{BRSAM}} & \multicolumn{2}{c|}{\textbf{MAM}} & \multicolumn{2}{c|}{\textbf{BRSAM}} \\
        \midrule
        \multicolumn{1}{|c|}{} & 
        \textbf{FPL-UE-A} & \textbf{HERDS} & \textbf{FPL-UE-A} & \textbf{HERDS} & \textbf{FPL-UE-A} & \textbf{HERDS} & \textbf{FPL-UE-A} & \textbf{HERDS} & \textbf{FPL-UE-A} & \textbf{HERDS} & \textbf{FPL-UE-A} & \textbf{HERDS} \\
        \midrule
        \textbf{3} & 0.975 & 0.077 & 0.241 & 0.097 & 0.178 & 0.079 & 0.375 & 0.214 & 0.135 & 0.051 & 0.385 & 0.314 \\
        \hline
        \textbf{4} & 0.988 & 0.083 & 0.347 & 0.153 & 0.242 & 0.099 & 0.417 & 0.342 & 0.179 & 0.055 & 0.524 & 0.428 \\
        \hline
        \textbf{5} & 0.110 & 0.101 & 0.452 & 0.245 & 0.271 & 0.116 & 0.665 & 0.528 & 0.220 & 0.078 & 0.672 & 0.518 \\
        \hline
        \textbf{6} & 0.237 & 0.122 & 0.457 & 0.269 & 0.307 & 0.132 & 0.687 & 0.574 & 0.351 & 0.087 & 0.735 & 0.588 \\
        \hline
        \textbf{7} & 0.248 & 0.146 & 0.555 & 0.321 & 0.351 & 0.138 & 0.674 & 0.592 & 0.342 & 0.089 & 0.745 & 0.594 \\
        \hline
        \textbf{8} & 0.254 & 0.148 & 0.561 & 0.375 & 0.434 & 0.156 & 0.757 & 0.621 & 0.401 & 0.100 & 0.801 & 0.621 \\
        \bottomrule
    \end{tabular}
    }
\end{table*}

\end{document}